\documentstyle[aps,preprint,prb]{revtex}
\draft

\begin{document}
% \tighten
\draft
% \sloppy

\title{Comment on `Magnetoresistance and differential
   conductance in mutliwalled carbon nanotubes'}
\author{Christian~Sch\"onenberger $^{\star}$ and Adrian~Bachtold $^{\ddagger}$}
\address{
   $^{\star}$Institut f\"ur Physik, Universit\"at Basel,
   Klingelbergstr.~82, CH-4056 Basel,
   Switzerland \\
   $^{\ddagger}$ Department of Physics,
   University of California and Materials Sciences Division,
   Lawrence Berkeley National Laboratory, Berkeley,
   CA~94720, USA}
\date{\today}
\maketitle

\begin{abstract}
Jeong-O Lee {\it el al.} [Phy.\ Rev.\ B, {\bf 61}, R16 362 (2000)]
reported magnetoresistance and differential
conductance measurements of multiwalled carbon nanotubes.
The observed aperiodic conductance fluctuations
and the negative magnetoresistance was interpreted to
originate exclusively from changes in the density of
states at
the Fermi energy. We show that
this interpretation is questionable and not supported
by their measurements.
\end{abstract}

\vspace{.5cm}
\vspace{.5cm}

In a recent paper Jeong-O Lee {\it el al.} discuss
measurements of the electrical resistance $R$ of
multiwalled carbon nanotubes (MWNTs).\cite{Lee}
In perpendicular magnetic field $H$ the resistance
decreases with field, i.e. displaying a
negative magnetoresistance (MR). In
addition, aperiodic resistance
fluctuations are superimposed.
The fluctuations and the
negative MR increase in magnitude at
lower temperature $T$. These charcteristic
features have been
seen before by several groups and were sucessfully
interpreted within the framework of quantum
interference
corrections to the diffusive motion of
electrons.\cite{Langer,Baxendale,BachtoldAPL,BachtoldNATURE,CS-ApplPhysA,naud}
In this interpretation the negative MR is caused by
weak localization, while the aperiodic conductance fluctuations
ressemble so-called
universial conductance fluctuations (UCF).
There is not only qualitative agreement between
previous measurements and
theory, but quantitative aggreement has been
obtained!\cite{BachtoldNATURE,CS-ApplPhysA}

As the authors mention, there is a disagreement between the
theoretical prediction and these previous experiments.
For a defect free (and undoped) metallic carbon nanotube
with ideal electrical contacts the
electrical conductance $G=1/R$ is predicted by theory to be {\em
twice} the quantum conductance $G_0=2e^2/h$ due to two propagating
one-dimensional ($1$d) modes
at the Fermi energy.\cite{Saito98}
This is not observed in experiments. Instead, $R$ is temperature
dependent, it increases if T is decreased. Also,
$R$ is strongly magnetic field-dependent, both in previous
experiments, as well as in the
experiments of  Lee {\it el al.}
However, for an ideal (and undoped) metallic carbon nanotube
the number of $1$d subbands at the Fermi energy
does not change if a perpendicular magnetic-
field
is applied. Hence, $R$ should be independent of
$H$. Any dependence on $H$ and $T$ for
temperatures below the $1$d-subband separation
points to physics which is beyond the simple and
extremely idealized picture of a
$1$d ballistic wire with zero
backscattering and non-interacting
electrons.
This should hold for the equilibrium (linear
response) resistance {\em and} for the
differential resistance, as long as the
applied voltage is lower than the $1$d-
subband separation.
There is only a disagreement between
theory and experiments if one sticks to the assumption
that nanotubes are ideal. This poses no problem to
theory, for which an ideal nanotube is the most simple
model to work with.
But why should a {\em real} nanotube be perfect
in the experiment? Nanotubes may have defects, adsorbates
may play a role, the evaporated metallic contacts
most likely add additional backscattering and
electrons in $1$d strongly interact. There is a large body of
experiments showing that all necessary ingredients for ballistic
transport are
not realized in  MWNTs.\cite{Langer,BachtoldNATURE,CS-ApplPhysA,BachtoldAFM}

Though Lee {\it el al.} measured MR dependences similar to
previously published work, they decided to explain their data along
a different line of thinking. They set out to prove that
the negative MR and the aperiodic fluctuations have nothing to do
with conventional interference corrections (WL, UCF), but mainly
originate from the change in the density-of-states (DOS) near the
Fermi energy
$E_F$. In order to support their statement, they not only
measure the equlibrium conductance but study the
differential
conductance $dI/dV(V)$ as a function of applied bias $V$,
too. Any change in $dI/dV$ is assumed to originate from a
change of the DOS. This interpretation is very problematic,
because
of the low-ohmic contacts to the nanotubes
and the four-terminal measurements.
Only in the opposite limit with {\em high-ohmic} contacts is
it possible to measure exclusively the DOS.
One has to make sure that the contacts (or at least one
contact) act as tunneling
contacts determining the total resistance locally.
Assume the contacts of Lee {\it el al.} were perfect, i.e. no
backscattering. Then, any change in $dI/dV$ would be due to
variations in the
transmission probability inside the nanotube. The DOS would
not matter at all, as long as the number of subbands is not
changed.

Lee {\it el al.} find in two MR measurements particular
field values at which the
measured resistance is practically temperature independent
(\mbox{$7$\, T} for sample S1 and \mbox{$4$\,T}
for sample S2). It is quite
interesting that the corresponding
resistance value
are close to the predicted value of
\mbox{$6.4$\,k$\Omega$} for a perfect
nanotube. However, this cannot be taken as a
proof for ballistic transport in the
nanotubes in agreement with the prediction
$G=G_0$ for an ideal tube. Following
the arguments of Lee {\it el al.}, we could equally well take
another data point of the  S2 data (see inset of Fig. 2 of Lee {\it
el al.}), where $R$ is also practically $T$ independent at
\mbox{$1.8$\,T}. According to Lee {\it el al.}
this would mean metallic (and ballistic) behaviour, this
time however with a resistance of \mbox{$7.2$\,k$\Omega$}
in contradiction with $G=2G_0$.

Furthermore, why no taking the
zero field data of Fig.~3 (of Lee {\it el al.}) which also
displays a temperature independent resistance $R$ below
\mbox{$4$\,K}. Following the reasoning of Lee {\it el al.},
the MWNT should also be metallic (and ballistic) at
\mbox{$H=0$\,T}. The disagreement
with $G=2G_0$ is now even larger.

Measuring $dI/dV$, Lee {\it el al.} have
observed pseudo-gaps of order \mbox{$1.5$\,mV} for certain
field values. The authors realize
that these gaps are an order of magnitude too small to be
explained
by theory (i.e. the separation between $1$d subbands of an ideal
nanotube). Moreover, they conclude their paper by mentioning
that `the most unusual observation is the existence
of apperiodic fluctuations of the
MR in perpendicular field totally absent in the theoretical
predictions'. The conclusion, which one should have drawn from
these
inconsistencies, is that the interpretation in
terms of DOS effects of an ideal nanotube is
wrong. The measurements cannot be explained by
simple DOS features obtained from a tight-binding
band-structure calculation.

In view of the author's own summary
we are quite irritated by the statement that
`the aperiodic
fluctuations and negative magnetoresistance mainly
originate from the change of density of states near the
Fermi level with
magnetic field, rather than a quantum interference effect',
which appears in the abstract. The authors provide no support
for this claim, they even have not tried to demonstrate that
their data cannot be understood in the framework of
quantum interference corrections.

In conclusion, the paper by Lee {\it el al.}
does not prove that the observed MR
in MWNTs is mainly due to DOS effects, it is rather in
support
of previous interpretations which
proved that interference corrections are important in
MWNTs.\cite{Langer,Baxendale,BachtoldAPL,BachtoldNATURE,CS-ApplPhysA,naud}

Finally, let us empasize that we do not claim that DOS effects
are unimportant in nanotubes at all. According to the Einstein
relation
the conductance is a product of the DOS and the diffusion coefficient
$D$. In the conventional theory of quantum corrections to the Drude
resistance, the main effect of interference is to change $D$, while
the interaction enters in to the DOS. This is only an approximation
valid for small corrections. Because corrections are large in MWNTs,
the two contributions
cannot easily be separated anymore.

\end{document}